\documentclass[11pt]{article}
\usepackage{moriond,epsfig}
\usepackage{pennames}

\def\Journal#1#2#3#4{{#1} {\bf #2}, #3 (#4)}

\def\PLB{{\em Phys. Lett.}  B}

\def\ZPC{{\em Z. Phys.} C}
\def\EPJC{{\em Eur. Phys. J.} C}
\def\CPC{\em Comp. Phys. Comm.}

\def\etal{{\it et~al.}}%

\def\eV{\ifmmode\mathrm{e\kern-0.08em V} \else e\kern-0.08em V\fi}%
\def\MeV{\ifmmode\mathrm{M}\else M\fi{\eV}}%
\def\GeV{\ifmmode\mathrm{G}\else G\fi{\eV}}%
\def\TeV{\ifmmode\mathrm{T}\else T\fi{\eV}}%
\newcommand{\LEP}{{\scshape lep}}
\newcommand{\ALEPH}{{\scshape aleph}}
\newcommand{\OPAL}{{\scshape opal}}
\newcommand{\DELPHI}{{\scshape delphi}}
\newcommand{\Lthree}{{\scshape l}{\small 3}}
\newcommand{\PYTHIA}{{\scshape pythia}}

\newcommand{\BEtt}{\ensuremath{{\mathrm{BE}_{32}}}}

\begin{document}
\vspace*{4cm}
\title{BOSE-EINSTEIN AND OTHER CORRELATIONS IN HADRONIC Z DECAY}

\author{ W.J. METZGER}

\address{University of Nijmegen, Toernooiveld 1, 6525\ ED\ \ Nijmegen, The Netherlands}

\maketitle\abstracts{
In hadronic Z decays, Bose-Einstein correlations in \Pgpz\ pairs
are compared to those in identical charged pion pairs.  
Bose-Einstein correlations are also measured
in triplets of identical charged pions, and comparison with those
in pairs of pions indicates that pion production is completely incoherent.
Further, factorial cumulants are used to compare correlations among
like-sign as well as
among all particles with those of several Monte Carlo models.
}

\section{3-particle BEC}
Bose-Einstein correlations (BEC) among $n$ identical particles
are usually studied in terms of 
\begin{equation}
    R_n(Q_n) = \frac{\rho_n(Q_n)}{\rho_{n0}(Q_n)}      \label{Rn}
\end{equation}
where $\rho_n(Q_n)$ is the $n$-particle number density as a function
of $Q_n$ and $\rho_{n0}$ is the density that would occur in the absence
of BEC.  
Here, 
$Q_2^2=M_{12}^2-4m_\pi^2$ is the four-momentum difference of a pair of particles,
and
$ Q_3^2={M_{123}^2-9m_\pi^2}=Q_{12}^2+Q_{23}^2+Q_{13}^2$ with
$Q_{ij}$ the four-momentum difference between the two pions $i$ and $j$ of the triplet.
In the simplest case of a Gaussian-shaped source of radius $R$, 
$R_2$ can be parametrized as
\begin{equation}   \label{R2param}
    R_2(Q) = {\cal N}(1+\alpha Q)(1+\lambda e^{-Q^2R^2})
\end{equation}
where ${\cal N}(1+\alpha Q)$ serves as normalization, 
taking into account some long-range correlation in $Q$.  
The density, $\rho_{n0}$, of the so-called reference sample is usually
taken from unlike-sign pairs, mixed events, or Monte Carlo.  
The use of unlike-sign pairs is problematic because of the presence of
many resonances.  The studies reported here do not use unlike-sign pairs
as reference sample.

The parameter, $\lambda$, is equal to unity if pion production is incoherent
and all pions come from the hypothesized source.  
The latter is certainly not the case because of long-lived resonances, 
and usually $\lambda$ is found to be less than unity. 
Whether production is indeed incoherent can be investigated by comparing
2- and 3-particle BEC.  This was done by \Lthree, and agreement with
the hypothesis of complete incoherence was observed.\cite{3p}
This is rather surprizing within a string picture of hadronization, since
particles produced close to each other along the string should be rather coherent.
The question is whether resonances would be sufficient to destroy this coherence.

\section{2-particle BEC: \boldmath{$\pi^0\pi^0$} and
                         \boldmath{$\pi^\pm\pi^\pm$}}
\begin{figure}[bht]
\begin{center}
\begin{minipage}{.49\textwidth}
  \includegraphics[width=.9\textwidth]{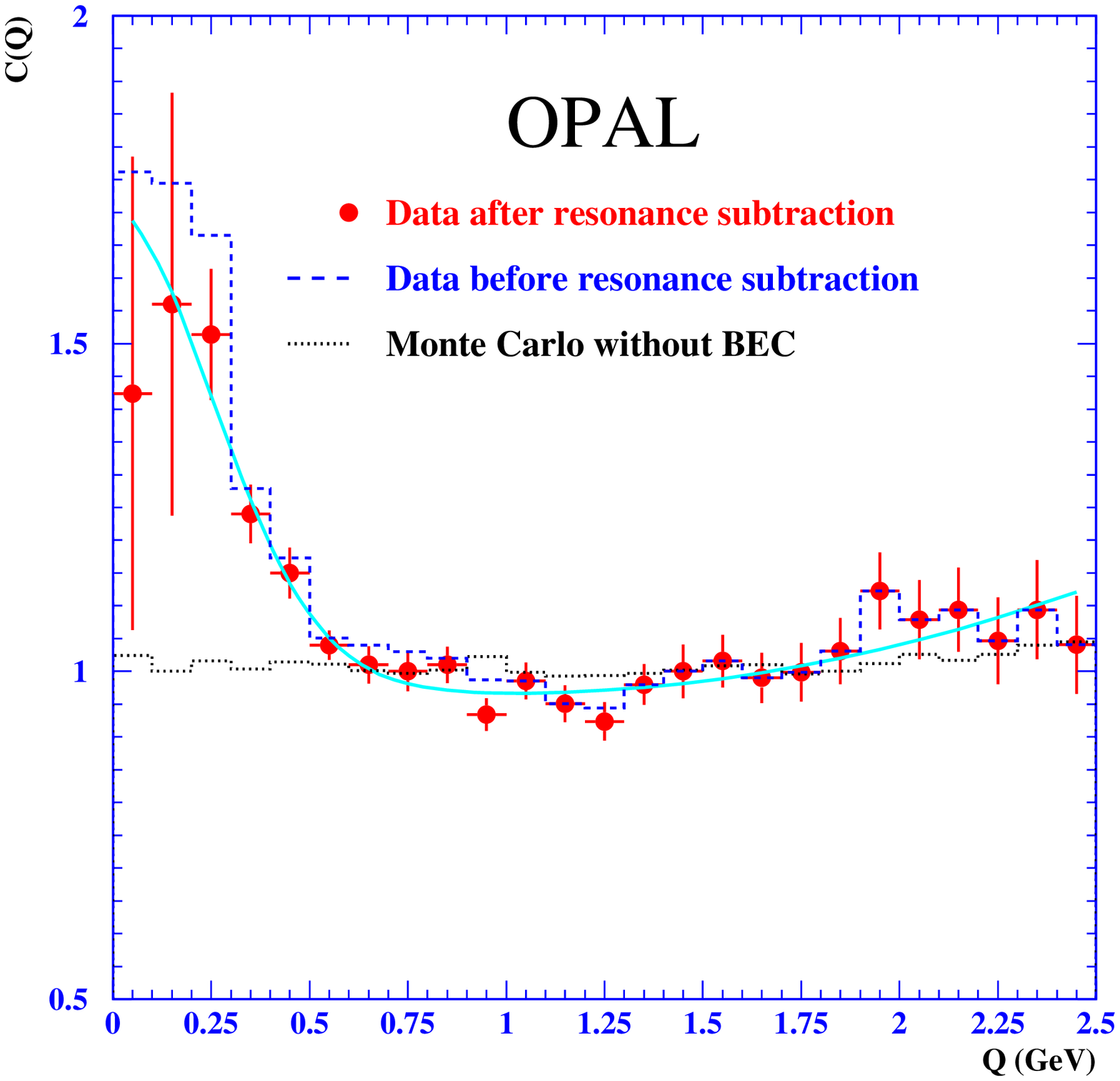}
\end{minipage}
\begin{minipage}{.41\textwidth}
 \begin{center}
  \includegraphics[width=.9\textwidth]{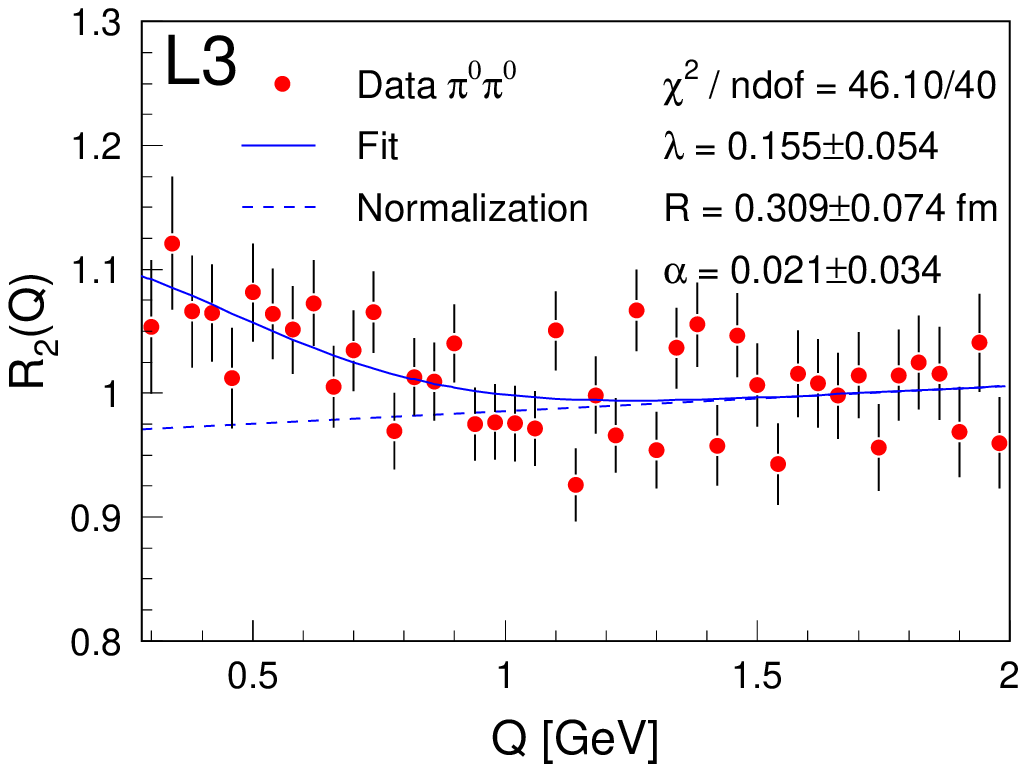}
  \includegraphics[width=.9\textwidth]{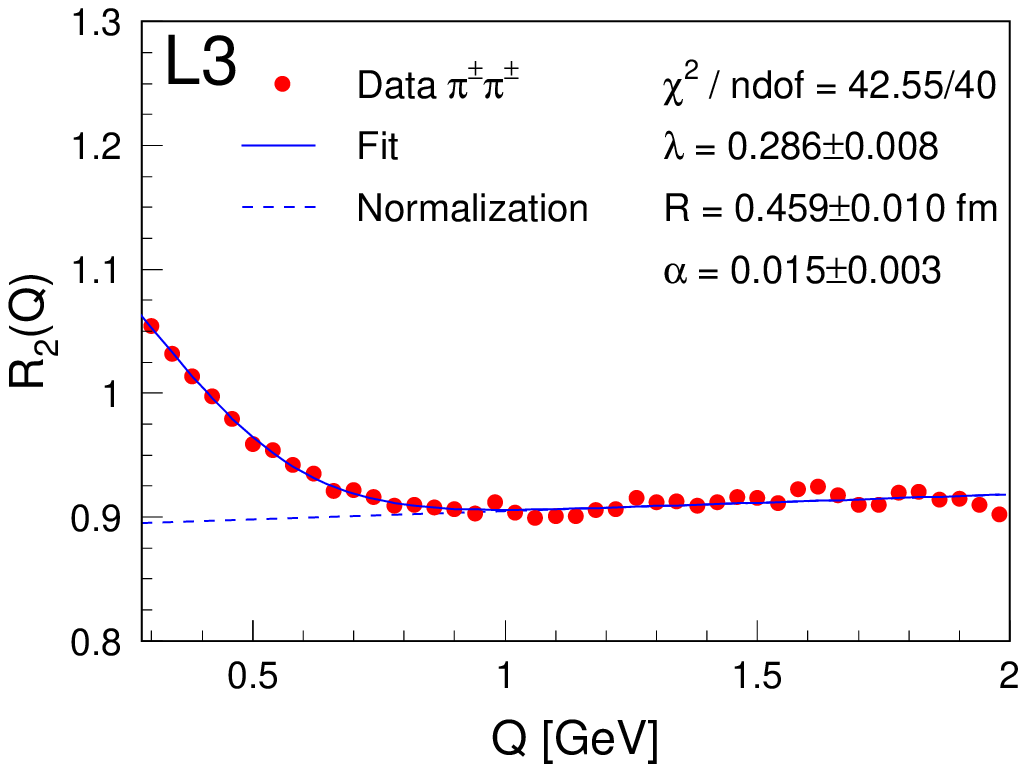}
 \end{center}
\end{minipage}
\caption{Distribution of $R_2$ for $\pi^0$ pairs in \OPAL\ and \Lthree,
as well as that for identical charged pion pairs in \Lthree\ subject to
selections analogous to the $\pi^0$ case.
}
\label{fig:Op}
\end{center}
\end{figure}

Measurement of BEC between neutral pions is quite rare, 
for obvious experimental reasons. In \Pep\Pem\ interactions this has only
been done by \Lthree\,\cite{2pL} and, more recently, by \OPAL.\cite{2pO}
Experimental exigencies result in
quite different analyses.  For example, \OPAL\ requires 2-jet events
(thrust, $T>0.9$) and $p_{\pi^0}>1\,\GeV$,
whereas \Lthree\ uses all topologies and requires $E_{\pi^0}<6\,\GeV$.
The results of both experiments are shown in Fig.~\ref{fig:Op}
and, together with those for charged pions from previous experiments,
in Table~\ref{tab:2p}.

\begin{table}[t]
\caption{Results of 2-particle BEC studies for charged 
and for neutral pions.\label{tab:2p}}
\vspace{0.4cm}
\begin{center}
  \begin{tabular}{|l|l|c|c|}  \hline
  Charges & Experiment                          &    $R$ (fm)   &   $\lambda$  \\ \hline
  $\pm\pm$& \ALEPH\,\cite{2pA}                    & $0.52\pm0.02$ & $0.30\pm0.01$ \\
          & \DELPHI\,\cite{2pD}                   & $0.47\pm0.03$ & $0.24\pm0.02$ \\
          & \OPAL\,\cite{2pOc}                    & $0.79\pm0.02$ & $0.58\pm0.01$ \\
          & \Lthree\,\cite{3p}                    & $0.65\pm0.04$ & $0.45\pm0.07$ \\
          & \Lthree\,\cite{3p} (3-\Pgp)           & $0.65\pm0.07$ &  $0.47\pm0.08$ \\
          & \Lthree\,\cite{2pL} $E_\Pgp<6\,\GeV$  & $0.46\pm0.01$ & $0.29\pm0.03$ \\ \hline
  00      & \Lthree\,\cite{2pL} $E_\Pgp<6\,\GeV$  & $0.31\pm0.10$ & $0.16\pm0.09$ \\
          &\OPAL\,\cite{2pO} $p_\Pgp>1\,\GeV$, 2-jet    & $0.59\pm0.09$ & $0.55\pm0.14$ \\ \hline
  \end{tabular}
\end{center}
\end{table}

The \Lthree\ values of 
both $\lambda$ and $R$ are about 1.5 standard deviations lower for
neutral than for charged pions.  Such a result would seem expected in
fragmentation models with local charge compensation, such as a string,
since two \Pgpz's could be produced next to each other while two \Pgpp's
could not.
However, resonances would dilute this effect.  
And \OPAL\ does not confirm it,\cite{2pO}
preferring to compare their \Pgpz\ result
to the \LEP\ average rather than 
to their own previous charged-$\pi$ result.\cite{2pOc}

In fact,
at first glance, the \Pgpz\ results of \Lthree\ and \OPAL\ disagree,
both for $\lambda$ and $R$, by about 2 standard deviations.
However, we note that the \Lthree\ values for charged pions are
lower when cuts are imposed analogous to those for the neutral pion
analysis.  The question is then whether the apparent \Lthree-\OPAL\
disagreement is not due to the different cuts applied in the analyses.
One hopes that \OPAL\ will do a charged-pion analysis 
using the same cuts as in their neutral-pion analysis.

\section{BEC and intermittency}
Factorial cumulants have been used by \OPAL\ to investigate correlations
among charged particles\,\cite{intall} and
among like-sign charged particles.\cite{intlike}
The factorial cumulant, $K_q$, of order $q$ measures the genuine correlations
among $q$ particles.

\begin{figure}[tbh]
\begin{center}
  \includegraphics[width=.44\textwidth]{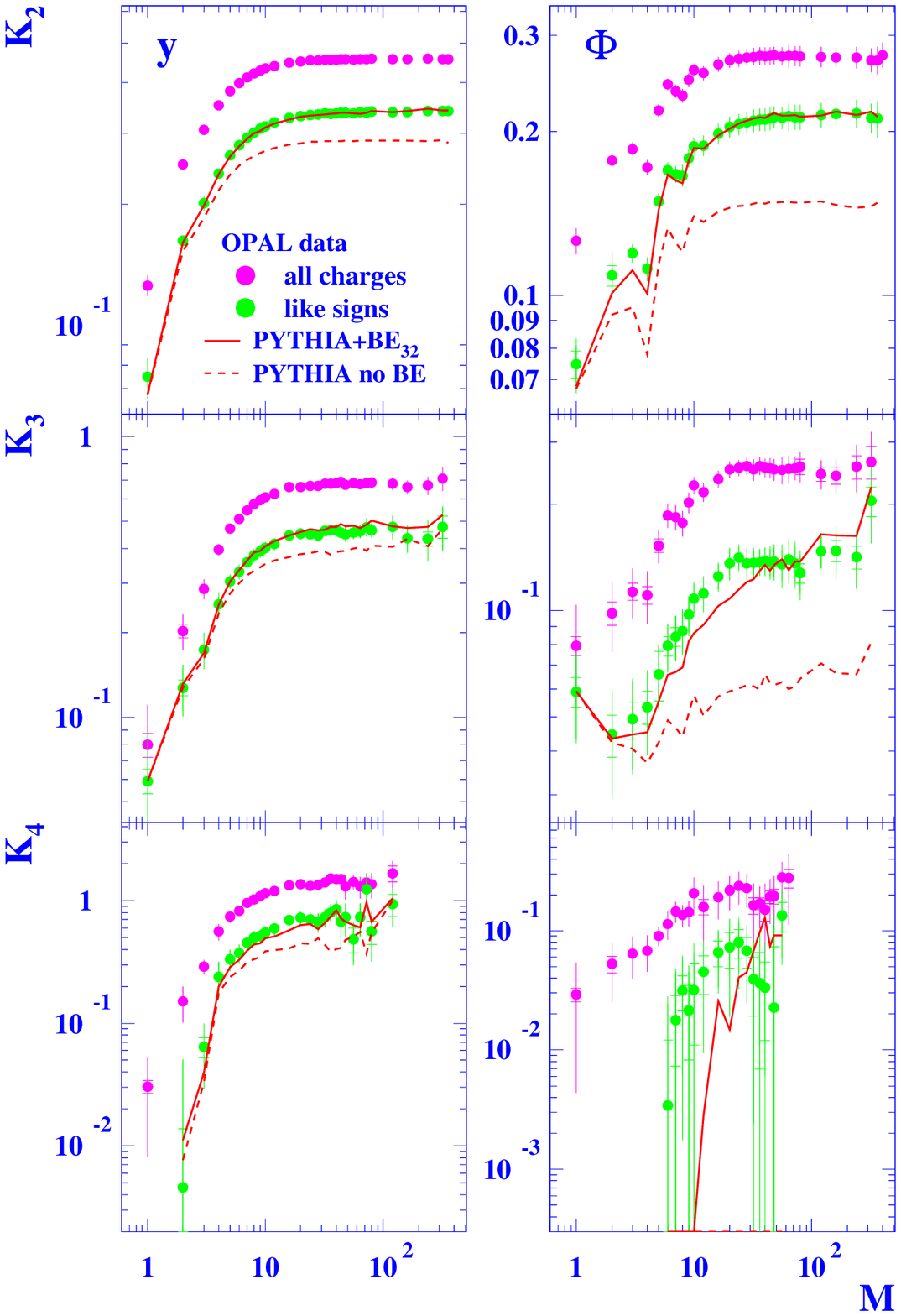}
  \includegraphics[width=.44\textwidth]{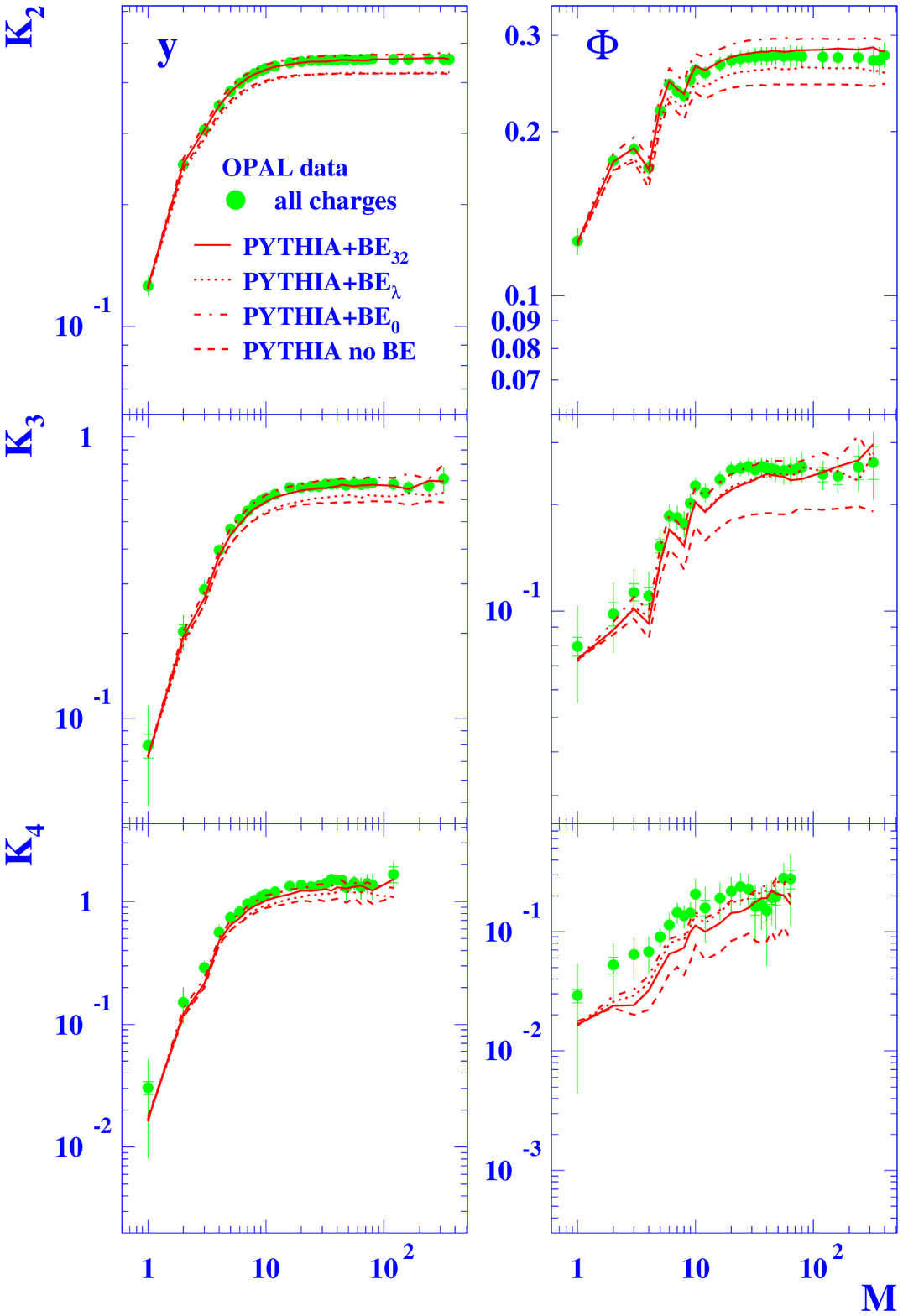}
\caption{Factorial cumulants compared to \PYTHIA\ Bose-Einstein models..
}
\label{fig:OK}
\end{center}
\end{figure}

The \OPAL\ results are shown in Fig.~\ref{fig:OK} for binning in rapidity, $y$,
and azimuthal angle, $\phi$, with respect to the sphericity axis.
The data are compared
to various Bose-Einstein models\,\cite{BEmod} available in \PYTHIA.\cite{pyth}

There is good agreement between the data and the \BEtt\ model,
while \PYTHIA\ without BE agrees less well.  
Quite remarkably, this is true not only for $K_2$ 
but also for $K_3$ and $K_4$ even though \BEtt\ contains no explicit
3- or 4-particle correlations.
Other BE models agree less well.

\begin{figure}[tbh]
\begin{center}
  \includegraphics[width=.32\textwidth]{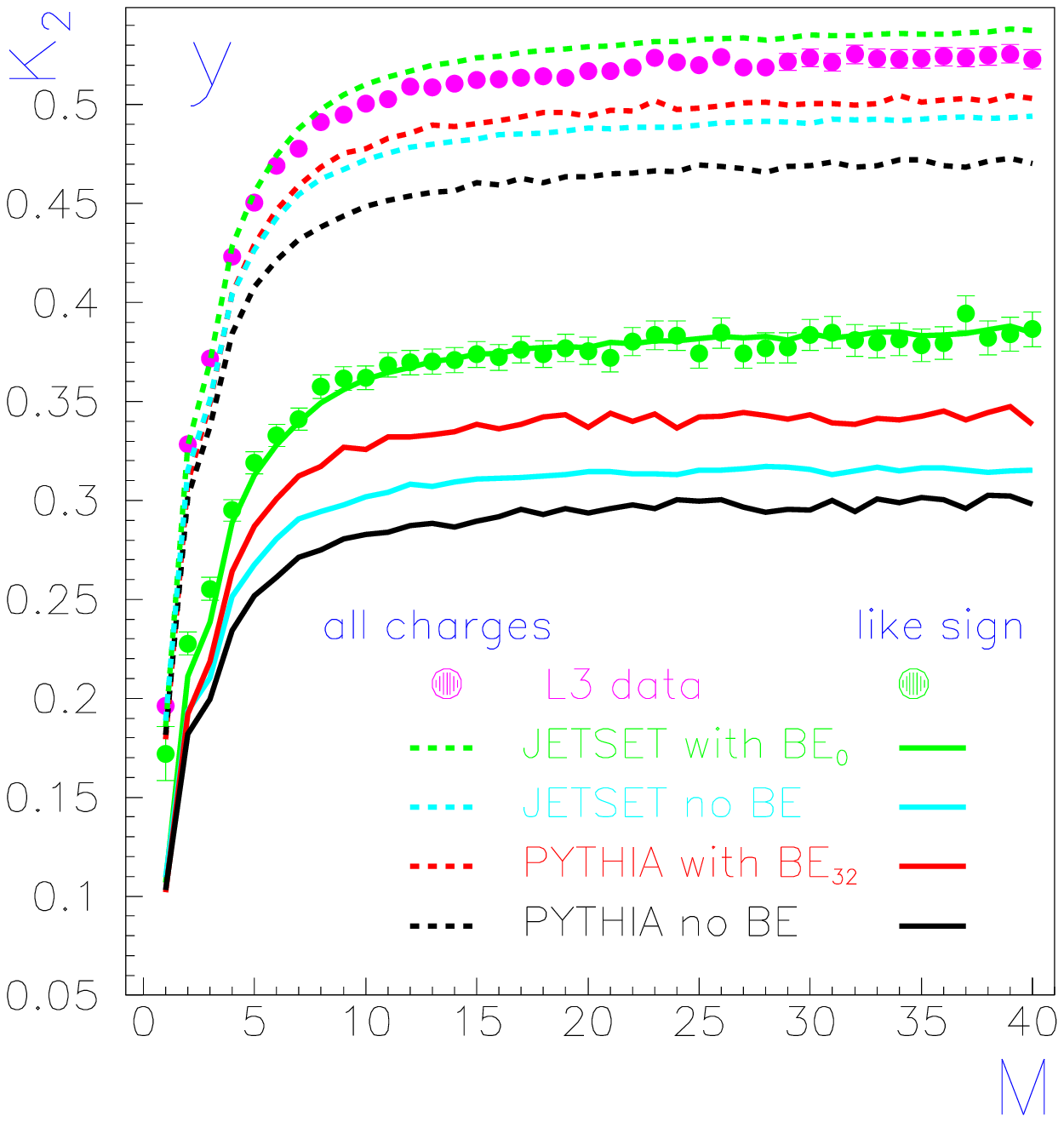}
  \includegraphics[width=.32\textwidth]{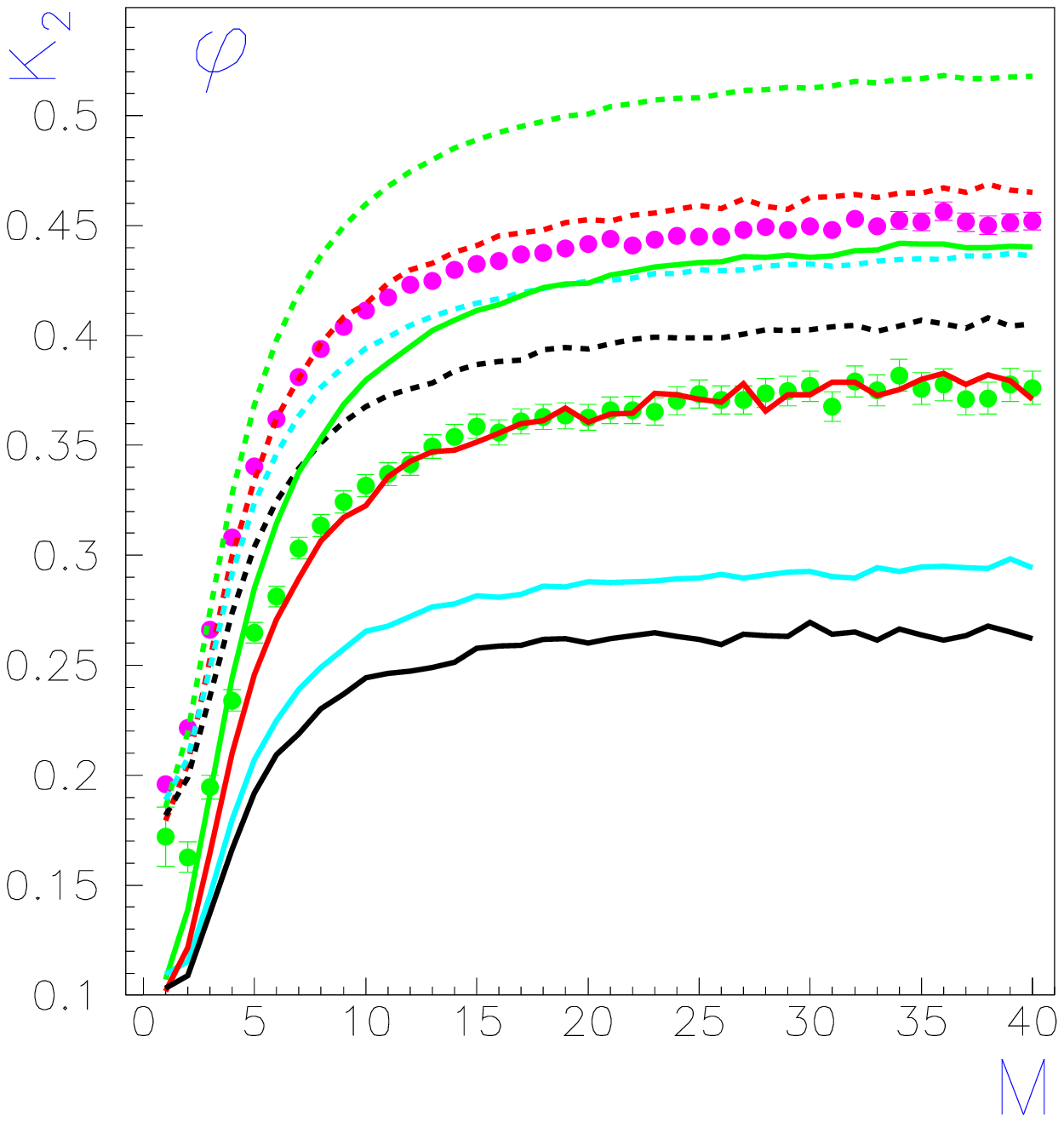}
  \includegraphics[width=.32\textwidth]{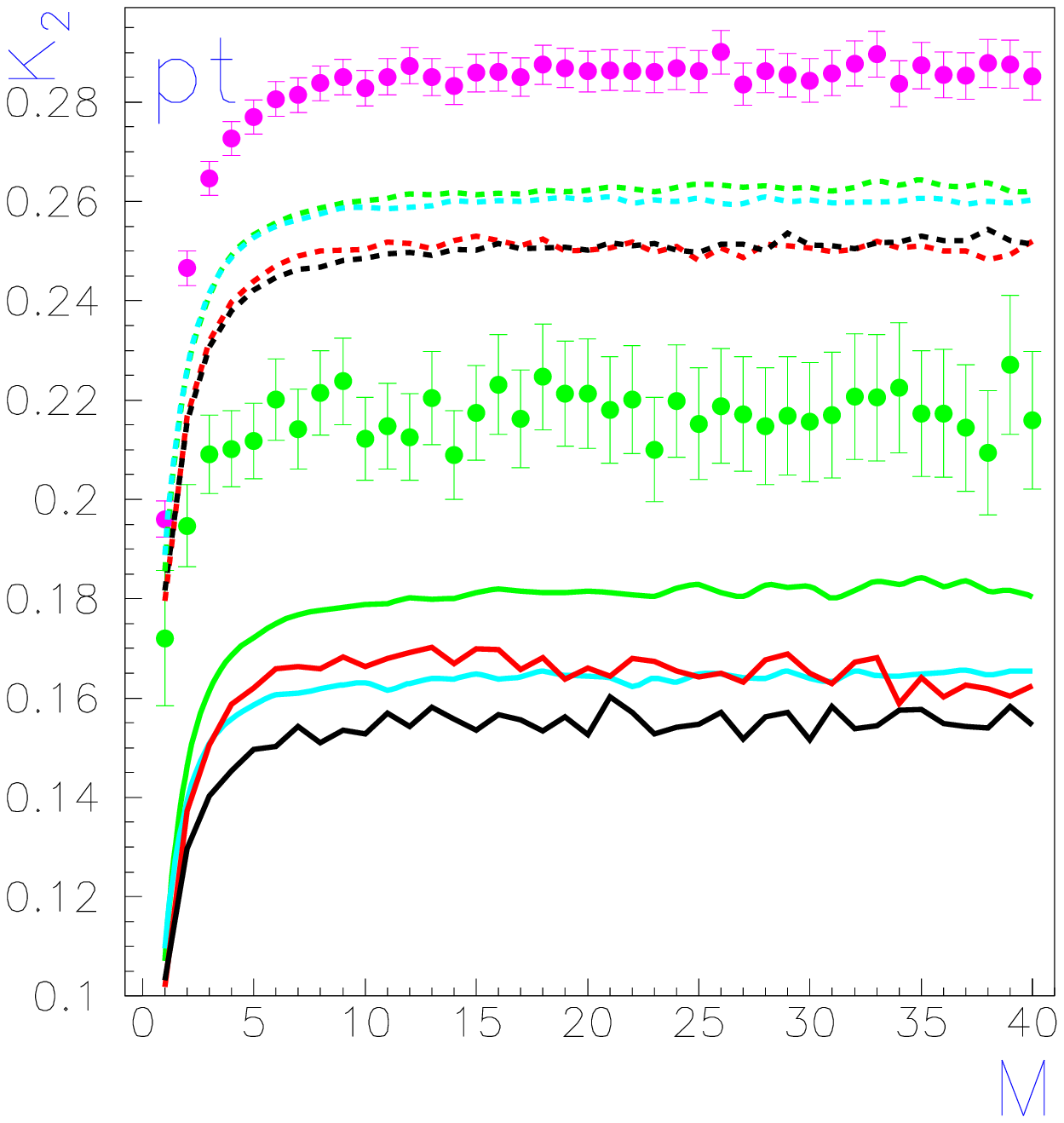}
  \includegraphics[width=.32\textwidth]{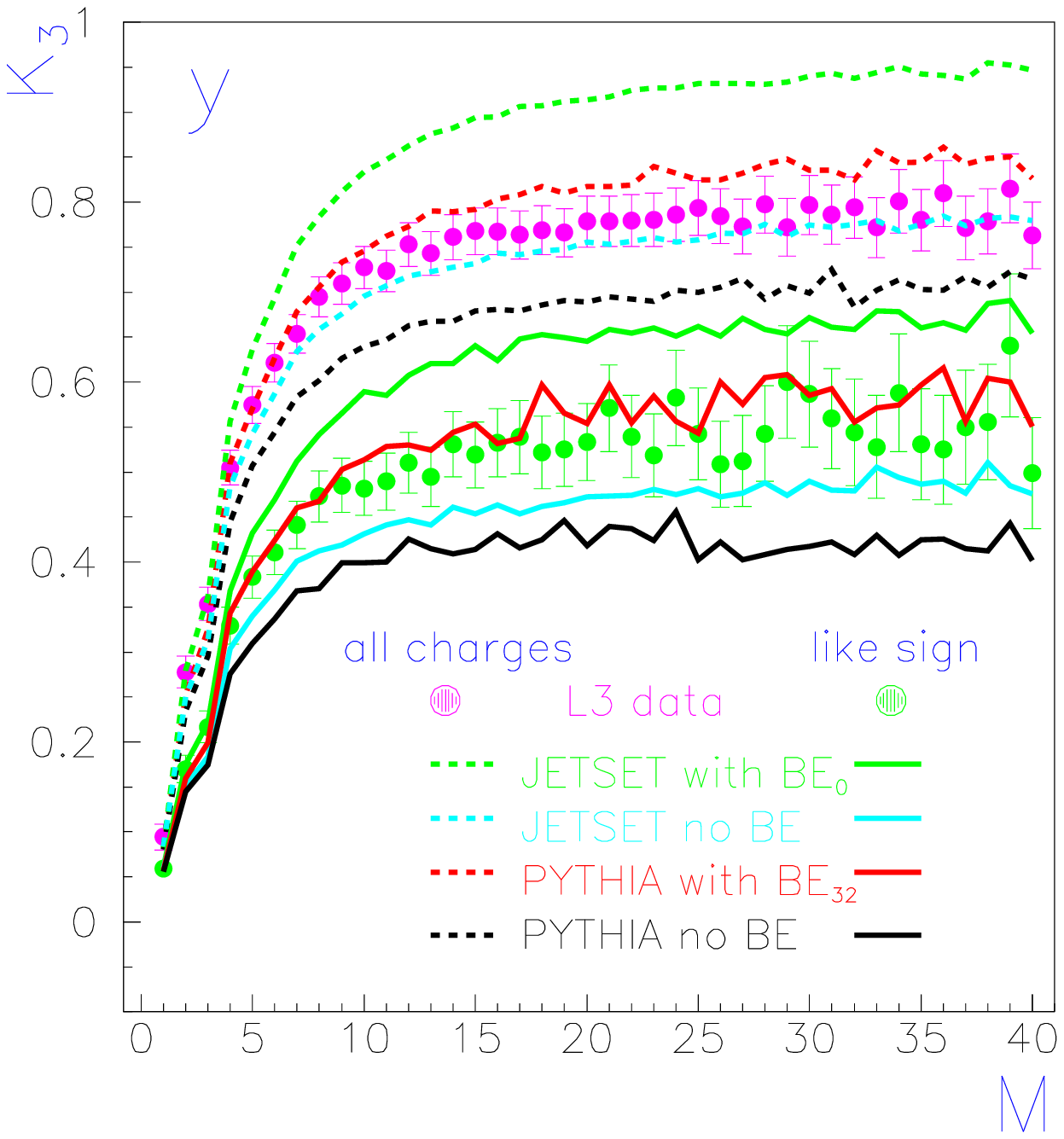}
  \includegraphics[width=.32\textwidth]{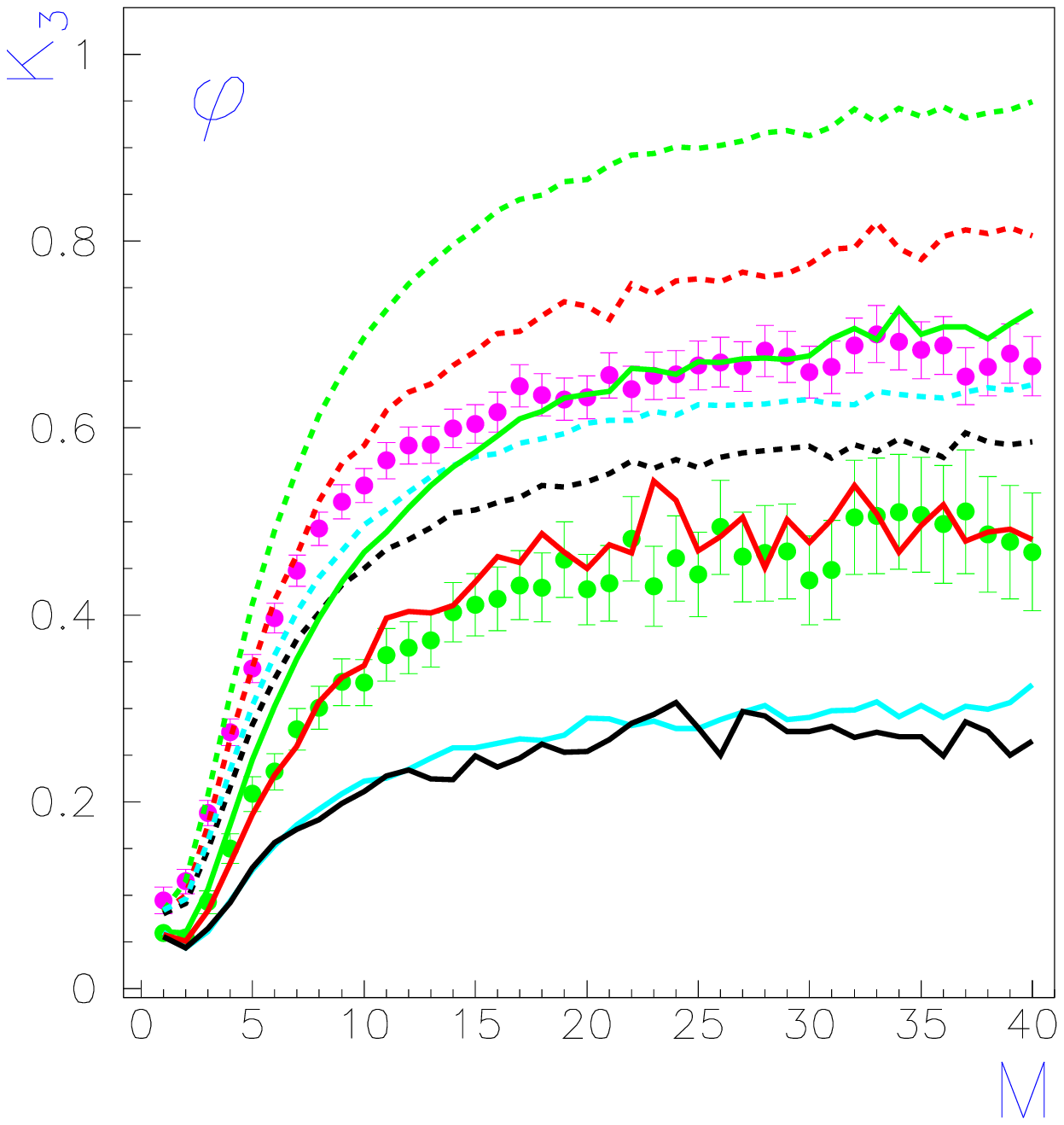}
  \includegraphics[width=.32\textwidth]{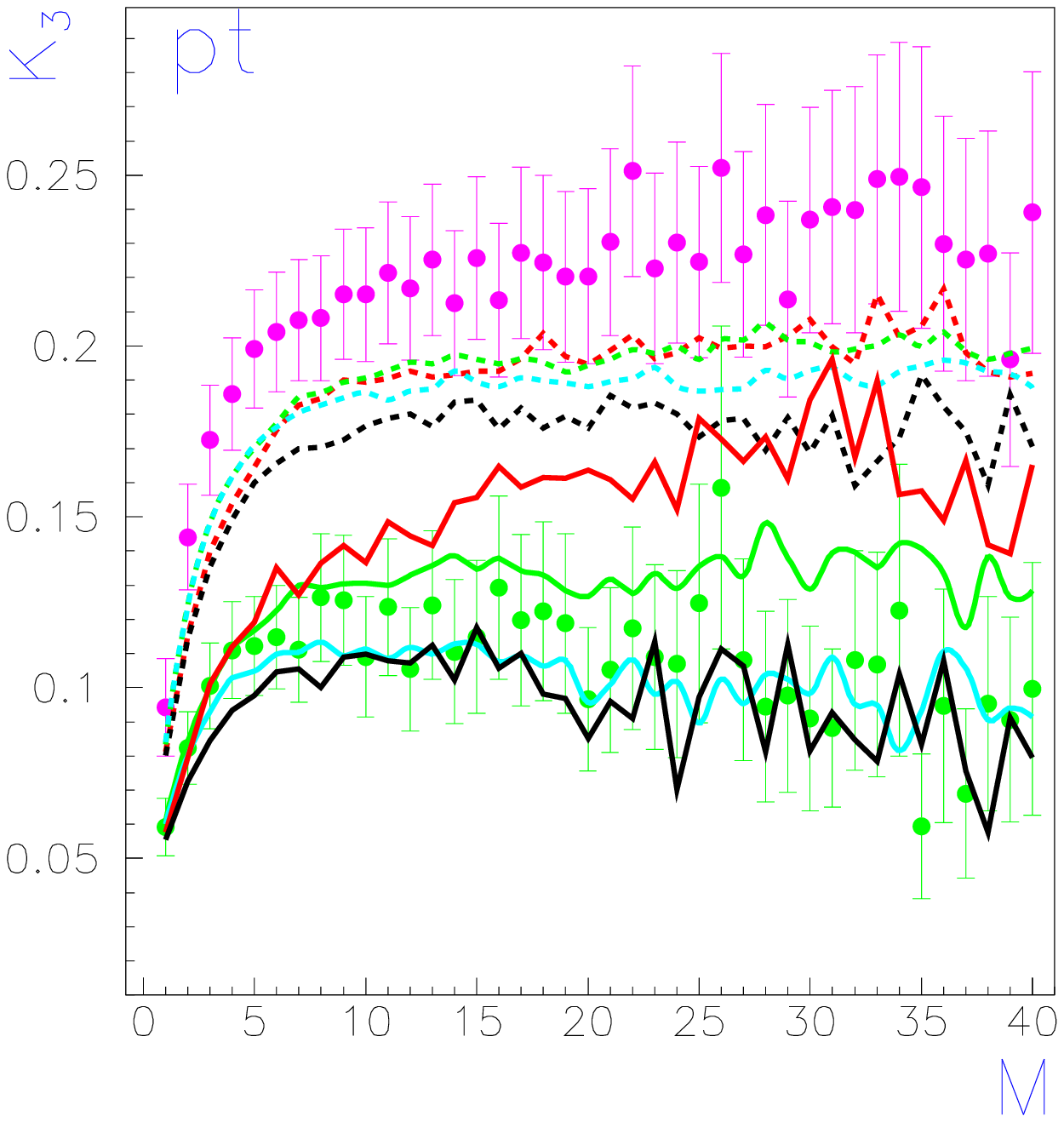}
\caption{Factorial cumulants compared to \PYTHIA\ Bose-Einstein models..
}
\label{fig:L3K}
\end{center}
\end{figure}

However, a recent preliminary \Lthree\ analysis 
with respect to the thrust axis, 
shown in Fig.~\ref{fig:L3K}, presents a less optimistic picture.
While \BEtt\ agrees better, overall, with the data, 
While overall agreement with the data is best for \BEtt,
the agreement for $K_2(y)$ cannot be deemed good, 
and none of the models gives good agreement in $\phi$.

It must be pointed out that these comparisons may be quite sensitive 
not only to the details and the values of parameters of the Bose-Einstein model, but also to
the fragmentation parameters of \PYTHIA.  The comparisons suggest, however, that it
might be useful to include these cumulants in the tuning of fragmentation
and BE-model parameters.

\section*{Acknowledgments}
I have benefited from discussions of the \OPAL\ results with
Dr.\,Edward Sarkisyan.  
I would also like to thank Ms.\,Qin~Wang, who performed the preliminary
\Lthree\ factorial cumulant analysis.

\end{document}